\documentclass[10pt,prl,twocolumn,showpacs]{revtex4-1}
\usepackage{graphicx,amsmath,amssymb,amsthm}
\usepackage{hyperref}
\usepackage{cleveref}
\usepackage{xcolor}
\usepackage{ulem}
\usepackage{thmtools, thm-restate}

\begin{document}
\newtheorem*{conjecture}{Distance Conjecture:}
\def\nn{\nonumber}
\def\kc#1{\left(#1\right)}
\def\kd#1{\left[#1\right]}
\def\ke#1{\left\{#1\right\}}
\newcommand\eqn[1]{\label{eq:#1}}
\newcommand\beq{\begin{equation}}
\newcommand\eeq{\end{equation}}
\newcommand\eq[1]{eq.~(\ref{eq:#1})}
\newcommand\tbl[1]{Table~\ref{tab:#1}}
\newcommand{\sn}[1]{\S~\ref{sec:#1}}
\newcommand{\fig}[1]{Fig.~\ref{fig:#1}}
\newcommand{\app}[1]{App.~\ref{app:#1}}
\renewcommand{\Re}{\mathop{\mathrm{Re}}}
\renewcommand{\Im}{\mathop{\mathrm{Im}}}
\renewcommand{\b}[1]{\mathbf{#1}}
\renewcommand{\c}[1]{\mathcal{#1}}
\renewcommand{\u}{\uparrow}
\renewcommand{\d}{\downarrow}
\newcommand{\be}{\begin{equation}}
\newcommand{\ee}{\end{equation}}
\newcommand{\bsigma}{\boldsymbol{\sigma}}
\newcommand{\blambda}{\boldsymbol{\lambda}}
\newcommand{\Tr}{\mathop{\mathrm{Tr}}}
\newcommand{\sgn}{\mathop{\mathrm{sgn}}}
\newcommand{\sech}{\mathop{\mathrm{sech}}}
\newcommand{\diag}{\mathop{\mathrm{diag}}}
\newcommand{\Pf}{\mathop{\mathrm{Pf}}}
\newcommand{\half}{{\textstyle\frac{1}{2}}}
\newcommand{\sh}{{\textstyle{\frac{1}{2}}}}
\newcommand{\ish}{{\textstyle{\frac{i}{2}}}}
\newcommand{\thf}{{\textstyle{\frac{3}{2}}}}
\newcommand{\SUN}{SU(\mathcal{N})}
\newcommand{\N}{\mathcal{N}}

\title{Distance Conjecture and De-Sitter Quantum Gravity}

\author{Hao Geng}
\affiliation{Department of Physics, University of Washington, Seattle, WA, 98195-1560, USA}

\date\today

\begin{abstract}
Motivated by the early discovery of the gigantic landscape of string theory vacua, in recent years people switched direction to try to find constraints on low energy effective field theories from UV-complete descriptions for example quantum gravity or string theory. Those constraints are called swampland criteria. Among the proposed swampland criteria the distance conjecture is the most well-motivated one from the universal behavior of string theory compactifications. It claims that for a type of scalar fields, called moduli, the effective field theory description will break down if it moves a large distance in field (moduli) space. In this work, we will show that in one proposal for a complete theory of quantum gravity including positive vacuum energy, the so called DS/dS correspondence, there is a strong signal of the validity of the distance conjecture.
\end{abstract}

\maketitle

{\bf Introduction:}
The seeking of unification in fundamental particle physics has been a long-term program starting from seminal works \cite{Glashow:1961tr,Weinberg:1967tq,Salam:1968rm} unifying electromagnetic and weak interactions. A unified description at high energy scale should be able to constrain the physics at low energy scale, for example the existence of neutrinos and their couplings to charged leptons from electroweak unification. String theory provides the only existing consistent framework to unify all known fundamental interactions including gravity. However, due to the large gap of string scale to electroweak scale and the large amount of vacua constructed in string theory \cite{Bousso:2000xa}, it is hard to study the implications of string theory to reality both experimentally and theoretically. Motivated by the idea of the existence of universal constraints on low energy physics from a unifying high energy theory, several swampland conjectures (for example \cite{Vafa:2005ui,Ooguri:2006in,Ooguri:2018wrx}) are proposed to distinguish those low energy effective theories that can be constructed from string theory or consistent with quantum gravity from those not. Among those conjectures, the distance conjecture \cite{Ooguri:2018wrx} is the most universal (for any spacetime) and best motivated one from ubiquitous behaviors of string theory compactification.
\begin{conjecture}
As the moduli move a distance beyond Plank scale in moduli space, a tower of light states emerges with masses exponentially suppressed.
\end{conjecture}
In this statement, moduli refer to scalar fields parametrizing the compactification and moduli space is the space describing the geometry of moduli fields defined by metric $g_{ab}(\phi)$ from the kinetic term of the Lagrangian density of moduli fields
\be
   \mathcal{L}_{\text{kinetic}}=\frac{1}{2}g^{ij}(\phi)\partial^{a}\phi_{i}\partial_{a}\phi_{j}.\label{eq:mm}
\ee
The prediction of this conjecture of low energy physics is that, for example, as a scalar field moves a large distance beyond the Plank scale, due to the tower of light states, the effective field theory description breaks down. Studying these arguments for low energy theories on de-Sitter space will be relevant to our real world.

One would like to base these arguments on a complete definition of string theory that encompasses more than just its perturbative or low energy corners (low energy effective field theory studies of some aspects of the distance conjecture can be found in \cite{Ooguri:2006in,Heidenreich:2018kpg,Geng:2019phi}). In its most concrete incarnation, string theory provides a non-perturbative definition of quantum gravity via holography \cite{Maldacena:1997re}, or the AdS/CFT correspondence, which equates quantum gravity in anti de-Sitter space with a standard field theory in one dimension less. Holography should be the gold standard from which to derive rigorous bounds on what can, and what can not, happen in quantum gravitational theories. The problem with this approach as it stands is that AdS/CFT as we understand it the best applies to {\it Anti} de-Sitter space, not de-Sitter space: the cosmological constant, and hence the vacuum energy, have to be negative. To address the distance conjecture in de-Sitter space, we need a holographic definition of quantum gravity in the presence of a positive vacuum energy. One proposal \footnote{Other proposals for holographic definitions of de-Sitter quantum gravity have been put forward, probably most notable the so called dS/CFT correspondence \protect{\cite{Strominger:2001pn}}. Both DS/dS and dS/CFT are \protect{``derived"} from standard AdS/CFT. In dS/CFT one attempts to basically Wick rotate AdS/CFT. This is clearly not how dS is realized in string theory - Wick rotation would require un-physical complex fluxes. Instead DS/dS is taking the basic physical insights of AdS/CFT as its starting point as described in the text.} for how to do this is the so called DS/dS correspondence \cite{Alishahiha:2004md}. This is the framework in which we will operate.

{\bf DS/dS and Entanglement:} Let us briefly review the key concepts behind the DS/dS correspondence. To understand the basic idea, it is useful to express the metric on a $D$ dimensional de-Sitter space of curvature radius $L$ as
\beq
\label{eq:warpedmetric}
ds^2_{DS_D} = dr^2 + e^{2A(r)} ds^2_{dS_{d}}
\eeq 
where $e^A = L \cos(r/L)$ and $d=D-1$. Note that the warpfactor $e^A$ vanishes linearly at the horizons at $r/L = \pm \pi/2$ and has a maximum at the central ``UV slice" at $r=0$. If we were to study AdS$_D$ instead the warpfactor would be a hyperbolic sine, also linearly vanishing at the horizon but growing without bound away from it. In this AdS$_D$ case the geometry has a dual description in terms of a CFT on dS$_d$, with the region near the horizon (where the warpfactor vanishes) being dual to the low energy (IR) degrees of freedom of the CFT. But near the horizon the two geometries (AdS$_{D}$ and DS$_{D}$) are indistinguishable. That motivated the authors of \cite{Alishahiha:2004md} to conjecture that the infrared degrees of freedom of the same CFT also holographically capture the near horizon degrees of freedom of DS$_D$. The fact that in the dS case the warpfactor reaches a maximum gets translated into imposition of a UV cutoff on the CFT. The fact that we see a second near horizon region beyond the central slice translates into a picture with two CFTs. Last but not least, building on the observation of \cite{Karch:2003em} that the geometry of DS$_D$ traps a massless $d$ dimensional graviton on the UV slice, it was conjectured that the holographic dual in the DS case contains dynamical gravity as well: DS$_D$ quantum gravity is holographically dual to two copies of a cut-off CFT in $d=D-1$ dimensions coupled via $d$ dimensional gravity. This prescription completely mirrors the holographic interpretation of the Randall-Sundrum braneworld scenario \cite{Randall:1999vf}. The only difference being that the latter requires us to introduce, by hand, a positive tension brane into AdS$_D$ to implement the UV cutoff and trap the dynamical graviton, whereas in the DS$_D$ case this happens automatically; the UV slice is an intrinsic part of the dS geometry without any additional matter. The most concrete implementation of this idea has been accomplished in \cite{Gorbenko:2018oov} where it was shown how to systematically start from a CFT dual to AdS$_D$ and go over to DS$_D$ by first adding a particular deformation that corresponds to removing the UV part of the geometry and then adding a different deformation that adds back in the UV of DS$_D$ instead.

Being rooted in the standard AdS/CFT correspondence, DS/dS inherits a prescription for how to calculate entanglement entropies in the $d$-dimensional field theory in terms of minimal surfaces in the $D$ dimensional geometry \cite{Ryu:2006bv} as was explored in \cite{Dong:2018cuv,Geng:2019bnn}. For a DS$_D$ spacetime these calculations allowed two very different ways to understand the Gibbons Hawking (GH) entropy associated with the de Sitter horizon. Here it doesn't matter whether we look at the $D$ or $d$ dimensional GH entropy; they are identical \cite{Karch:2003em}. The calculations of \cite{Dong:2018cuv} determined the entanglement between the two CFTs of the holographic dual on the entire spatial slice of the dS$_d$ spacetime they live on, finding perfect agreement with the DS$_D$ entropy. We'll refer to this entanglement between the two CFTs as the "left/right entanglement". In \cite{Geng:2019bnn} it was found that the same numerical value of the entanglement entropy also governs the entanglement between degrees of freedom within the dS$_d$ static patch (the region of dS$_d$ accessible to a single observer) and those behind the horizon. Last but not least, one can calculate the left/right entanglement between the two CFTs, but restricted to the degrees of freedom in the static patch. The left/right entanglement is extensive, so restricting to the volume within the static patch simply yields half the GH entropy since we only have half the dS spatial volume within the static patch. Nevertheless, this quantity will play an important role in what follows.

This picture changes dramatically if we add extra interaction terms to the field theory Lagrangian. In the standard spirit of holography adding a scalar operator to the Lagrangian corresponds to turning on a non-trivial profile for a scalar field. In \cite{Geng:2019bnn} we insisted that the bulk metric still takes the form of \eqref{eq:warpedmetric}, supported by a scalar profile $\phi(r)$. Note that these solutions preserve the full isometry of the dS$_d$ space on each slice. We take this to mean that we are still studying the vacuum state of a field theory on dS$_d$. In DS/dS the full $D$ dimensional symmetry is emergent to begin with from the field theory point of view. By adding extra couplings to the $d$ dimensional field theory we ruin this symmetry enhancement to $D$ dimensional dS isometries. But by insisting on $d$ dimensional dS isometries we are guaranteed to still study ground states. In the presence of these scalar deformations the two entanglement entropies above start differing from each other. The entanglement of degrees of freedom across the dS$_d$ horizon continues to agree perfectly with the $d$-dimensional GH entropy for arbitrary warpfactor $A(r)$; the same integral of $A$ determines modifications to the bulk minimal area as well as the boundary Newton's constant, modifying GH entropy and entanglement entropy in exactly the same way. The left/right entanglement on the other hand starts to disagree with the GH entropy. In fact, it was shown in \cite{Geng:2019bnn} that the left/right entanglement entropy on the entire spatial slice of dS$_d$ is always {\it larger} than the dS$_d$ GH entropy, at least as long as the extra matter that is responsible for the deformation obeys the null energy condition. A strong interpretation of the dS$_d$ GH entropy is that it represents the maximum dimension of the Hilbert space in which all excitations within the dS$_d$ static patch live. If we adopt this point of view our result indicates that in order to properly account for the full left/right entanglement we do need to access degrees of freedom behind the horizon. A single dS$_d$ observer could never reconstruct the left/right entangling surface in the bulk, and hence could never reconstruct the entire bulk geometry.

The situation is even more dramatic if the bulk deformation forces even the left-right entanglement {\it within} the dS$_d$ static patch to be larger than the GH entropy. Recall that in this case we started out with a factor of 2 separating the two in the case of undeformed DS$_D$. If we once again adopt the strong interpretation that the GH entropy represents a limit on the size of the Hilbert space accessible to degrees of freedom within the static patch, we seem to be led to a clear contradiction. Deformations with this property would give an entanglement entropy for states within this Hilbert space that exceeds its dimension. Theories for which this happens should then be banned to the swampland. What we left unanswered in our previous work is under what circumstances this happens. In this letter, with a careful identification of the moduli space distance \footnote{A recent definition of ``distance'' from the geometric point of view using Ricci flow is \protect{\cite{Kehagias:2019akr}}}, we will derive a bound on it, for a special type of matter deformation, from the requirement that a theory should not allow entanglement entropies for degrees of freedom entirely contained within the dS$_d$ static patch to exceed the dS$_d$ GH entropy. Then we will argue that, from this calculation, there is always an upper bound on the moduli distance for generic matter deformations preserving the dS structure. In the end we will discuss the possible future implications of inflationary cosmology models and its difficulty in our approach \footnote{Recent studies of the constraints on slow roll parameters and the dS swampland conjecture \cite{Obied:2018sgi,Garg:2018reu} using the so called transplankian censorship conjecture \cite{Bedroya:2019snp} includes \cite{Bedroya:2019tba,Brahma:2019vpl}.}.

{\bf Bound on Moduli Distance:} As derived in our previous work \cite{Geng:2019bnn} and summarized in the appendix, the ratio on left/right entanglement within the static patch to the GH entropy is given by
\beq R_S \equiv  \frac{S_{L/R}}{S_{GH}} = \frac{e^{(D-2) A(0)}L \int_{-\pi/2}^{0} d \beta 
\cos^{D-3} \beta}{\int_{r_{min}}^{r_{max}} e^{(D-3) A} dr }. \label{eq:ratio}\eeq
Here we take the UV slice at which $e^A$ takes its maximum value to be, without loss of generality, at $r=0$. $r_{min/max}$ correspond to horizons at which $e^A$ vanishes. Given our discussion above, we assume that a theory is in the swampland if it gives rise to $R_S > 1$. As we show in the appendix, at $D=4$, when we turn on a massless free scalar field this bound is equivalent to
\beq
\label{eq:bound}
   \delta s_{\text{in moduli space}}\leq 0.446 \frac{1}{\sqrt{G}}
\eeq
where $G$ is the 4-dimensional Newton constant which is related to Plank scale by $G=\frac{1}{M_{P}^2}$. The interpretation of this distance is as follows. This distance (see \eqref{eq:distance}) parametrized by bulk radial coordinate measures the distance between the solutions with arbitrary $A(0)$ and that with $A(0)=A_{\text{horizon}}=-\infty$. Because if $A(0)$ is exactly $A_{\text{horizon}}$ then the spacetime collapses and so is trivial. Hence we can choose that trivial geometry as the starting point in moduli space and going from the trivial solution to the solution with a given $A(0)$ in moduli space is equivalent to going in spacetime, with that given $A(0)$, from the horizon to the central slice. Here we emphasis that, besides the scalar field $\phi$, $A(0)$ is another modulus measuring the size of spacetime geometry which is specific in the context of de-Sitter quantum gravity. 

Beyond free massless scalar field, the bound \eqref{eq:bound} tells us that for general matter deformation the moduli distance has a component with an upper bound. This component is the projection of the moduli space trajectory onto the $A(0)$ axis. Hence for any reasonable trajectory (non-degenerate along the $A(0)$ axis) in moduli space the length should have an upper bound. And the bound \eqref{eq:bound} is understood to be an example that we can find the bound explicitly.   

{\bf Discussion and Future Remarks:} In this work, we show a strong signal that distance conjecture is valid in de-Sitter quantum gravity by an entropic consideration. The interesting thing is that, even though we did not directly address that the break down of effective field theory is due to the emergence of a tower of light states once the modulus moves a large distance, the entropy constraint can be thought of as a version the bound for the number of species of the degrees of freedom and hence equivalently claims that as the moduli fields moving the number of degrees of freedom is emerging and there is an upper bound of the number of degrees of freedom due to the existence of horizon in de-Sitter space. In a short sentence, distance conjecture preserves the consistency of de-Sitter structure. More of this important picture will be studied in \cite{Geng:2019phi}.

Moreover, from a more practical point of view, it will be interesting to see how much our holographic bound \eqref{eq:bound} can tell us about the constraints on cosmological models. For example, the constraints on slow roll parameters of inflation \footnote{H.Geng, to appear}. However, one of the difficulties is that, in our set-up, all the profiles are radial coordinate dependent instead of time dependent as that happened in inflationary models.
\section*{Acknowledgements} This work was supported in part by a grant from the Simons Foundation (651440, AK). I appreciate Andreas Karch, Dieter L$\ddot{u}$st and Matthew Reece for useful discussions. I am very grateful to my parents and recommenders.

\appendix
\section{Review of Previous Work}
{\bf Entropies:} RT formula for the D-type entangling surface provides:
\begin{equation}
    \begin{split}
        S_{L/R}&=\frac{A_{D}}{4G}\\
        &=\frac{L^{D-2}A_{D-3}e^{(D-2)A(0)}\int_{-\frac{\pi }{2}}^{0}d\beta\cos^{D-3}{\beta}}{4G}\\
        &=\frac{L^{D-2}A_{D-3}e^{(D-2)A(0)}\int_{-\frac{\pi }{2}}^{0}d\beta\cos^{D-3}{\beta}}{4g\int^{r_m}_{-r_m}e^{(D-3)A}dr}
    \end{split}
\end{equation}
The d-dimensional Gibbons-Hawking entropy is:
\begin{equation}
    \begin{split}
        S_{GH}=\frac{L^{D-3}A_{D-3}}{4g}
    \end{split}
\end{equation}
{\bf Conventions and Equations:} We have Einstein's equations and scalar equation of motion in spacetime \eqref{eq:warpedmetric}:
\begin{equation}
\begin{split}
\label{eq:Einstein1}
        8\pi GT^{r}_{r}&=-\frac{e^{-2A}}{2}\frac{(D-1)(D-2)}{L^{2}}+\frac{(D-1)(D-2)}{2L^{2}}\\
        &+\frac{(D-1)(D-2)}{2}A'^{2}
        \end{split}
\end{equation}
\begin{equation}
    \begin{split}
        8\pi G T^{t}_{t}&=(D-2)A''+\frac{(D-1)(D-2)}{2}A'^{2}\\
       & -e^{-2A}\frac{(D-3)(D-2)}{2L^{2}}+\frac{(D-1)(D-2)}{2L^{2}}
    \end{split}
\end{equation}
\begin{equation}
\label{eq:eom}
    \begin{split}
        \frac{d^{2}}{dr^{2}}\phi+(D-1)A'\frac{d}{dr}\phi-V'(\phi)=0
    \end{split}
\end{equation}
This tells us:
\begin{equation}
    \begin{split}
        A'(r)=\pm\sqrt{\frac{16\pi GT^{r}_{r}}{(D-1)(D-2)}-\frac{1}{L^{2}}+\frac{e^{-2A}}{L^{2}}}\label{eq:A'}
    \end{split}
\end{equation}
The warped factor integral is:
\begin{equation}
    \begin{split}
        \int_{-r_{m}}^{r_{m}}e^{(D-3)A}dr=2Le^{(D-2)A(0)}\int_{-\infty}^{0}\frac{e^{(D-3)\Delta}d\Delta}{\sqrt{e^{-2\Delta}-\frac{H(r)-1}{H(0)-1}}}\label{eq:warpedIntegral}
    \end{split}
\end{equation}
where we defined:
\begin{equation}
\label{eq:convenience}
    \begin{split}
        H(r)=\frac{16\pi G L^{2}}{(D-1)(D-2)}T^{r}_{r},\text{ and }\Delta=A(r)-A(0)
    \end{split}
\end{equation}
The scalar action is:
\begin{equation}
    \begin{split}
        S_{matter}=\int d^{D}x\sqrt{-g}(-\frac{1}{2}g^{ab}\partial_{a}\phi\partial_{b}\phi-V(\phi))
    \end{split}
\end{equation}
Then the energy momentum tensor reads:
\begin{equation}
        \begin{split}
        T_{ab}=-2\frac{1}{\sqrt{-g}}\frac{\delta S}{\delta g^{ab}}
        =-g_{ab}[\frac{1}{2}\partial_{c}\phi\partial^{c}\phi+V]+\partial_{a}\phi\partial_{b}\phi
    \end{split}\label{eq:EMT}
\end{equation}
\section{Deriving the Bound in D=4}
Turning on a free massless scalar field, the equation of motion \eqref{eq:eom} has the solution
\beq
 \phi'=Ce^{-3A(r)}.
\ee
Notice that this solution has an interesting behavior that if we approach the horizon where $A\rightarrow-\infty$ whereby $\phi'\rightarrow\infty$. This behavior is physically reasonable because horizon is the causal boundary and hence we are like trapped by an infinite potential well inside the horizon then $\phi'\rightarrow\infty$ means that we are never able to approach the horizon because that takes an infinite amount of energy. Using \eqref{eq:A'}, \eqref{eq:EMT} and the condition $A'(0)=0$, we can eliminate the constant $C$ by $A(0)$ through
\beq
    A'(0)^{2}=\frac{8\pi GC^2}{6}e^{-6A(0)}-\frac{1}{L^{2}}+\frac{e^{-2A(0)}}{L^{2}}.
\eeq
Then we use \eqref{eq:warpedIntegral} and \eqref{eq:ratio} to get
\beq
  R_{S}=\frac{e^{A(0)}}{\int_{0}^{1}dx\frac{1}{\sqrt{(x^{-2}-1)[1+x^{-2}(1+x^{-2})(1-e^{-2A(0)})]}}}\label{eq:changev}
\eeq
where we change variable from $\Delta$ to $x=e^{\Delta}$. $R_{S}$ as a function of $A(0)$ is monotonically increasing as can be seen numerically. As a result, the out of swampland criterion $R_{S}\leq1$ is equivalent to
\beq
  A(0)\leq 0.2037.\label{eq:newform}
\eeq
Now we can start to consider the distance moved in moduli space as we go from central slice to horizon
\beq
\begin{split}
  \delta s_{\text{in moduli space}}&=\int dr \sqrt{e^{3A}\frac{\phi'^2}{2}+\frac{e^{3A}}{8\pi G}6A'^{2}}\\&=M_{P}\sqrt{\frac{3}{8\pi}}
  \int_{0}^{1}dx x^{\frac{1}{2}}\times\\&\sqrt{\frac{3x^{-4}(1-e^{-2A(0)})-x^{2}+e^{-2A(0)}}{x^{-4}(1-e^{-2A(0)})-x^{2}+e^{-2A(0)}}}.\label{eq:distance}
  \end{split}
\eeq
where the Plank scale $M_{p}$ is just $\frac{1}{\sqrt{G}}$ and we used the same trick as \eqref{eq:changev} to change the integration variable from the bulk radial coordinate to the warped factor. As can be seen numerically that $\delta\phi$ as a function of $A(0)$ is monotonically increasing and so the out of swampland criterion \eqref{eq:newform} is finally equivalent to
\beq
  \delta s_{\text{in moduli space}}\leq 0.446 M_{p}.
\eeq
One thing deserves to be noticed is that since the bulk radial direction is a compact direction, its size is also a modulus and it can be characterized by $A(0)$.

\bibliography{Distance}

\end{document}